\documentstyle[12pt,psbox]{article}

\newcommand{\ginga}{{\it GINGA}\ }
\newcommand{\etal}{{\it et al.\ }}
\newcommand{\solarmass}{{$M_\odot$}}

\newcommand{\apj}{{\it Astrophys. J.}}
\newcommand{\aaa}{{\it A. \& A.}}
\newcommand{\mnras}{{\it Mon. Not. Roy. Astr. Soc.}}
\newcommand{\pasj}{{\it Publ. Astr. Soc. Japan}\ }
\newcommand{\nature}{{\it Nature}\ }

\begin{document}

\begin{flushleft}
\vspace*{1.7cm}
{\Large \bf CYCLOTRON LINE VARIABILITY}

\vspace{1.0cm}
\normalsize
T. Mihara$^1$,\ \ K. Makishima$^2$,\ \ and\ \ F. Nagase$^3$

\vspace{1.0cm}
{\it $^1$The Institute of Physical and Chemical Research,
2-1 Hirosawa, Wako, Saitama 351-01, Japan}\\
{\it $^2$Dept.\ of Physics, School of Science, Univ.\ of Tokyo,
7-3-1 Hongo, Bunkyo-ku, Tokyo 113, Japan}\\
{\it $^3$Inst.\ of Space and Astronautical Science,
3-1-1 Yoshinodai, Sagamihara, Kanagawa 229, Japan}
\end{flushleft}

\vspace{0.6cm}

\section*{\normalsize ABSTRACT}{
We systematically analyzed the spectra of X-ray binary pulsars observed with {\it GINGA} (Mihara 1995).
A new model NPEX (Negative and Positive power-laws EXponential) was introduced to represent the pulsar continuum.
Combining the NPEX continuum with the CYAB factor (cyclotron
resonance scattering model),
we successfully fit the whole-band spectra of all the pulsars.
A possible physical meaning of the NPEX model is the Comptonized spectra. 

\vspace{5.5mm} \hspace{-8mm} 
By using the smooth and concave NPEX model,
the cyclotron structures were detected from 12
pulsars, about a half of the 23 sources,
including new discoveries
from \mbox{LMC X-4} and \mbox{GS 1843+00}.
The magnetic fields were scattered in the range of $3\times10^{11}$ --
$5\times10^{12}$ G.  The distribution was shown for the first time,
which is  remarkably similar to that of radio pulsars
with a peak at $2 \times 10^{12}$ G.

\vspace{5.5mm} \hspace{-8mm} 
The double harmonic cyclotron structures of 4U 0115+63 in 1990  changed to a single structure in 1991.
The resonance energy also increased by 40 \% as the luminosity decreased to 1/6.
If we attribute this change to the height of the scattering region in a dipole magnetic field,
the height change is $\sim$ 1.1 km.
Such changes of the resonance energies  with luminosities are observed from  5
pulsars and can be explained by the accretion column height model.

\section*{\normalsize INTRODUCTION}

The X-ray binary pulsar is a neutron star in a contact binary system with mostly a high-mass star.
The neutron star is highly magnetized ($\sim10^{12}$G), collimates the
accreting matter onto magnetic poles and shows X-ray pulses with the rotation.
The electron cyclotron resonance structure in the X-ray spectrum is the only direct method to measure the
magnetic fields on the neutron star.
The resonance energy is $E_a [\mbox{keV}] = 11.6 B [10^{12}\mbox{G}]$.
The first report of the structure was from Her X-1 (Tr{\"u}mper \etal 1978),
followed by that from 4U 0115+63 by Wheaton \etal (1979).
But for the other pulsars, the magnetic fields are estimated by a
rather uncertain method using accretion spin up/down theory.
Spectra of X-ray binary pulsars look non-thermal and are not explained well.
Theories have been proposed (eg. Meszaros 1992), but
comparison with the data has not been done much.

\vspace{5.5mm} \hspace{-8mm} 
\ginga LAC (Turner \etal 1989) not only discovered cyclotron structures from many pulsars, but also
enabled us to discuss on the continuum spectra.
Those two are related with each other.
A good representation of the continuum spectra is essential to a
precise analysis of the cyclotron structure.
\ginga observed 23 pulsars including 1 pulsar-candidate with good statistics 
in its 4 years and 9 months life.
First we introduce a new empirical continuum model NPEX,  and discuss the meaning.
Next we show the magnetic fields distribution.
Last we discuss the variability of the cyclotron structure.

\section*{\normalsize NPEX MODEL}

It is known that a typical spectrum of a binary X-ray pulsar is a power-law (POWL) below $\sim$20 keV
and falls off exponentially at higher energies.
The reason of the exponential cutoff (ECUT) was not known.
\ginga found cyclotron structures at the bottom of the fall-off, which lead to a physical idea
that the ECUT is created by the cyclotron resonances of the fundamental, 2nd, 3rd,...  harmonics.
Thus early studies of pulsars with \ginga were done by employing a power-law (POWL) model as a continuum
and a cyclotron feature as an absorption (CYAB) (Makishima and Mihara 1992).
They employed two resonance, the fundamental and the 2nd harmonics, because those two are within the \ginga energy range.
This model succeeded to explain the overall spectrum in 8--60 keV of Her X-1 (Mihara \etal 1990).
It favored an absorption at 34 keV rather than an emission at 50keV which was uncertain in the previous observations.
\[
	CYAB(E) = e^{-\tau_1}, \hspace{1cm}
	\tau_1 = \frac{D_1 (\frac{W}{E_a}E)^2}{(E-E_a)^2 + W^2}, \hspace{1cm}
	\tau_2 = \frac{D_2 (\frac{2W}{2E_a}E)^2}{(E-2E_a)^2 + (2W)^2} ,
\]
Here $\tau_1$ is optical depth of the fundamental cyclotron scattering in a
classical cold plasma. $\tau_2$ is that for the 2nd harmonic.
$E_a$ is the resonance energy, $W$ is the width of the resonance, 
and $D_1$ and $D_2$ are the depths of the resonances.
Resonance energy and the width of the 2nd harmonic were fixed to the double of those of the fundamental in the fitting,
because the 2nd resonance was almost at the end of the energy range and it was difficult to be obtained independently.

\vspace{5.5mm} \hspace{-8mm} 
The flux of POWL$\times$double CYAB's model goes back to the POWL level far above the resonance.
Later the {\it HEAO-1} A4 spectrum of Her X-1 in 13--180 keV was published by Soong \etal (1990),
but the data does not show the flux return.
Putting the 3rd, 4th and 5th harmonics can reduce the flux, but it is not favorable because
it requires larger cross section of the 3rd than the 2nd,
larger the 4th than the 3rd.
It is possible that optical depth of the 2nd is apparently  larger than that of the fundamental
because the two-photon decay of the 2nd harmonic may fill up the fundamental,
but for higher harmonics than the 2nd, the reverse of the optical depths would not happen.

\vspace{5.5mm} \hspace{-8mm} 
Another problem is on 4U0115+63, which is the only pulsar with a clear 2nd harmonic observed with {\it GINGA}.
POWL$\times$ double CYAB's cannot explain the spectrum. The continuum needs to fall off by itself
(Nagase \etal 1991).

\vspace{5.5mm} \hspace{-8mm} 
In order to solve those problems, it is a better and natural idea to assume that
the continuum falls off  thermally by itself.
We tried some continuum models together with a single CYAB to the Her X-1 spectrum which has the best statistics.
We started with the simplest $\exp(-E/kT) \times$CYAB, but failed.
Next we tried Boltzmann model $E^{\alpha}\exp(-E/kT) \times$CYAB, which was successful in 13-60 keV.
But $\alpha$ became positive ($\alpha=0.74$) and cannot fit the negative POWL region below 10 keV.
Then, by adding negative POWL,  we introduce the NPEX (Negative and Positive power-laws EXponential) model  as
\[
  NPEX(E) = (A_1 E^{-\alpha_1} + A_2 E^{+\alpha_2} ) \ \exp \biggl(-\frac{E}{kT} \biggr)  ,
\]
where $kT$ is a typical temperature of the X-ray emitting plasma, and
$\alpha_1$ and $\alpha_2$ are the negative and positive POWL indices,
respectively.

\vspace{5.5mm} \hspace{-8mm} 
The NPEX$\times$CYAB model can fit the Her X-1 spectrum very well
in the entire 2--60 keV energy band with an iron line included.
Moreover the positive index converged to $\alpha_2  = 1.97 \pm 0.26$,
which suggests the blackbody ($\alpha_2  = 2$).
This model can fit the pulse-phase-resolved spectra, too.

\vspace{2mm}
\begin{table}[bht]
\small
\caption[]{
The best-fit parameters with NPEX$\times$CYAB model for the pulse
averaged spectra.
Errors are in 90\% confidence level.
Positive POWL index $\alpha_2$ is fixed to 2.0.
The units of $A_1$ and $A_2$ are [photons/s/keV/4000cm$^2$] at 10 keV.
$kT$, $E_{a}$, $W$ and $E_{\rm Fe}$ are in [keV],  $N_H$ is in [cm$^{-2}$], 
and $I_{\rm Fe}$ is in [photons/s/4000cm$^2$].
}
\label{NPEX para}
\begin{center}
\begin{tabular}{ccccccccccc}
\hline \hline
sources         & \multicolumn{2}{c}{Negative POWL}     & Pos. POWL      &   Exponential  & Absorption      \\
                &  $A_1$        &  $\alpha_1$           &     $A_2$      &    $kT$        & log$_{10} N_H $ \\
\hline
Her X-1        &$  135 \pm   8 $&$   0.51 \pm  0.03 $&$  100 \pm  23  $&$   8.0 \pm   0.8 $&$  $     ---    $ $\\
4U0115+63 (90) &$  491 \pm 326 $&$   0.41 \pm  0.48 $&$ 4960 \pm 1220 $&$   4.2 \pm   0.1 $&$  $     ---    $ $\\
4U0115+63 (91) &$   62 \pm  22 $&$   0.65 \pm  0.29 $&$  785 \pm  58  $&$   4.3 \pm   0.1 $&$  $     ---    $ $\\
X0331+53       &$  930 \pm  63 $&$  -0.27 \pm  0.05 $&$  630 \pm 170  $&$   6.3 \pm   0.5 $&$  $     ---    $ $\\
1E2259+586     &$   13 \pm  12 $&$   1.42 \pm  0.47 $&$    9 \pm   7  $&$   2.1 $ fixed    &$  $     ---    $ $\\
LMC X-4        &$   21 \pm   1 $&$   0.43 \pm  0.06 $&$   19 \pm   2  $&$   7.3 \pm   0.3 $&$  $     ---    $ $\\
GS1843+00      &$   45 \pm   3 $&$   0.73 \pm  0.08 $&$   47 \pm   5  $&$   8.2 \pm   0.2 $&$ 22.29 \pm  0.05 $\\
Cep X-4        &$  101 \pm  13 $&$   0.70 \pm  0.05 $&$  110 \pm  59  $&$   6.4 \pm   1.5 $&$ 22.01 \pm  0.08 $\\
Vela X-1       &$  171 \pm   4 $&$   0.61 \pm  0.05 $&$  123 \pm   8  $&$   6.4 \pm   0.1 $&$ 22.41 \pm  0.06 $\\
4U1907+09      &$   11 \pm   2 $&$   1.39 \pm  0.29 $&$   25 \pm   5  $&$   6.4 \pm   0.7 $&$ 22.86 \pm  0.08 $\\
4U1538-52      &$   19 \pm   2 $&$   1.47 \pm  0.20 $&$   68 \pm   8  $&$   4.6 \pm   0.2 $&$ 22.80 \pm  0.07 $\\
GX301-2        &$  135 \pm  71 $&$   0.80 \pm  0.85 $&$  485 \pm 184  $&$   5.4 \pm   0.3 $&$ 23.37 \pm  0.07 $\\
               & \multicolumn{3}{c}{Leaky absorber, Norm $ \times 0.38 \pm 0.26 $ }  &$   $&$ 24.44 \pm  0.22 $\\
\hline
\end{tabular}

\vspace{5mm}

\begin{tabular}{ccccccccccc}
\hline \hline
sources        &     & Resonance  & Width                      &        Depth         & Iron Flux & Energy  & $\chi^2_\nu$ \\
               &     & $E_{a}$    & $W$                        &        $D$           & $I_{\rm Fe}$  & $E_{\rm Fe}$  & \\
\hline							                               
Her X-1        &     &$  33.1 \pm   0.3 $&$  12.1 \pm   1.7   $&$   1.53 \pm   0.25 $ & 30$\pm$4       & 6.65 fix      & 1.14 \\
4U0115+63 (90) & 1st &$  11.3 \pm   0.6 $&$   5.9 \pm   0.8   $&$   0.67 \pm   0.08 $ & 17$\pm$18      & 6.60 fix      & 0.69 \\
               & 2nd &$  22.1 \pm   0.4 $&$   5.2 \pm   1.0   $&$   0.51 \pm   0.07 $ &                                \\
4U0115+63 (91) &     &$  15.6 \pm   0.4 $&$   9.0 \pm   0.6   $&$   1.22 \pm   0.06 $ &  ---           & ---           & 1.72 \\
X0331+53       &     &$  27.2 \pm   0.3 $&$   7.5 \pm   0.9   $&$   1.62 \pm   0.15 $ & 38 $\pm$ 16    & 6.59 fix      & 1.65 \\
1E2259+586     &     &$   4.2 \pm   0.6 $&$   2.0 \pm   0.9   $&$   0.86 \pm   0.27 $ &  ---           & ---           & 8.71 \\
LMC X-4        &     &$  21.4 \pm   1.2 $&$   5.1 \pm   3.8   $&$   0.11 \pm   0.05 $ & 2.7$\pm$0.4    & 6.6 $\pm$ 0.1 & 0.83\\
GS1843+00      &     &$  19.8 \pm   2.1 $&$   9.9 \pm   3.6   $&$   0.16 \pm   0.05 $ & 7.6$\pm$0.9    & 6.40 fix      & 0.49 \\
Cep X-4        &     &$  28.8 \pm   0.4 $&$  12.1 \pm   3.1   $&$   1.67 \pm   0.59 $ & 7 $\pm$ 2      & 6.5 $\pm$ 0.1 & 0.97 \\
Vela X-1       & 1st &$  24.5 \pm   0.5 $&$   2.2 \pm   1.0   $&$   0.065 \pm  0.015$ & 27 $\pm$ 3     & 6.5 $\pm$ 0.1 & 0.56 \\
               & 2nd &  $2 E_{a1}$ fixed &  $2 W_{1}$ fixed    &$   0.80 \pm   0.26 $ &                               \\
4U1907+09      &     &$  18.9 \pm   0.7 $&$   7.4 \pm   2.1   $&$   0.87 \pm   0.21 $ & 1.7 $\pm$ 0.5  & 6.60 fix      & 1.18 \\
4U1538-52      &     &$  20.6 \pm   0.2 $&$   4.2 \pm   0.6   $&$   0.83 \pm   0.08 $ & 2.5 fixed      & 6.50 fix      & 1.58 \\
GX301-2        &     &$  37.6 \pm   1.1 $&$  16.4 $ fixed      &$   0.65 \pm   0.17 $ & 32.1$\pm$ 3.4  & 6.60 fix      & 0.98 \\
\hline 
\end{tabular}
\end{center}
\vspace{-5mm}

\normalsize
\end{table}
  

\vspace{5.5mm} \hspace{-8mm} 
We applied this model to other pulsars with $\alpha_2$ fixed to 2.
Only NPEX continuum is used to those without a cyclotron structure,
NPEX$\times$CYAB is used to those with a single structure,
and NPEX$\times$double CYAB's is used to those with two harmonics,
which are  4U 0115+63 in 1990 and probably in Vela X-1.
A merit of NPEX continuum model is that it is slightly concave as is often seen in the pulsar spectra in 2--10 keV range.
With this continuum we discovered cyclotron structures from LMC X-4 and GS1843+00.
The fitting parameters are summarized in Table 1.

\section*{\normalsize MEANING OF NPEX MODEL}
Let us consider the physical interpretation of the NPEX model.
It would be natural to assume that $kT$ is the typical temperature of the X-ray emitting plasma.
We normalize the spectra with the energy of $kT$ after correcting the detector efficiency and the
absorption by the intervening matter.
The flux level is normalized by the flux at $E = kT$ (Figure 1).

\vspace{5.5mm} \hspace{-8mm} 
In the  case of  the non-cyclotron sources,
the spectra obey a power law in the low energies, but  with   different indices,
and show a round shoulder at around $E = 3 kT$.
Those are  represented by the negative and positive  POWL's of the NPEX model, respectively.
As the slope of the power-law flattens, the hump at $E = 3 kT$ increases,
suggesting the existence of ONE hidden parameter
which determines the shape of the continuum.
The hidden parameter is also suggested from the pulse profiles sliced by some energy bands.
Since 
the pulse shapes do not change below $kT$ and above $kT$,
the two POWL's cannot be independent, but are coupled by a hidden parameter.

\vspace{5.5mm} \hspace{-8mm} 
In the case of the cyclotron sources, 
the overall curves  are similar to those of the non-cyclotron sources.
The difference is that the spectrum shows
a steep fall-off at an energy, 
which is caused by the cyclotron resonance,
and in some pulsars reaches a local minimum at the resonance center.

\vspace{5.5mm} \hspace{-8mm} 
What mechanism creates both the negative POWL and the blackbody ?
The multi-blackbody model would be possible, but an artificial distribution of temperature is needed.
A better candidate is the Comptonization model.
In fact the changes of the spectra in Figure 1 reminds us the Comptonized spectra for different optical depth $\tau$.
If a soft photon goes into the hot electron plasma, where scattering is more dominant than absorption,
the photon gains energy by the inverse-Comptonization and
comes out with a larger energy.
When $\tau$ is small, the spectrum is a 
power law, and when $\tau$ is large, the Wien peak appears.
$\tau$ can be the hidden parameter.

\vspace{5.5mm} \hspace{-8mm} 
An analytic approximate calculation was done by Sunyaev and Titarchuk (1980) for a given soft photon input.
The emergent photon spectrum is generally given as $(E^2 + o(E^2)) \exp(-E/kT)$,
where $o(E^2)$ is the polynomials with lower order than 2.
For example, when the input photon has an index of $\alpha=-2$ in the high energy wing,
the output photon spectrum $F(x)$ is expressed as
\[
F(x) \ \propto \ e^{-x} \biggl( \frac{x^2}{24} + \frac{x}{6} + \frac{1}{2} + \frac{1}{x} + \frac{1}{x^2}\biggr), 
\hspace{1cm} x \equiv E/(kT).
\]
$x^2$ term corresponds to the positive POWL and
the rest is combined to a negative POWL.

\vspace{5.5mm} \hspace{-8mm} 
Figure 1 bottom is another Comptonization model by Lamb and Sanford (1979), (CMPL).
It assumes a bremsstrahlung as the input photon.
The change of the spectral shape with $\tau$ mimics that of the observed spectra.
Thus, Comptonization model is a very possible candidate.
The CMPL fits to the data   ($\chi^2_\nu = 1\sim3$) are not as good as NPEX fits, but it represents the overall shapes well.
It would be because the averaged spectrum cannot be represented by an ideal model.

\newpage
\begin{figure}[ht]
\begin{flushleft}
\psbox[xsize=0.375#1,ysize=0.375#1,rotate=r]{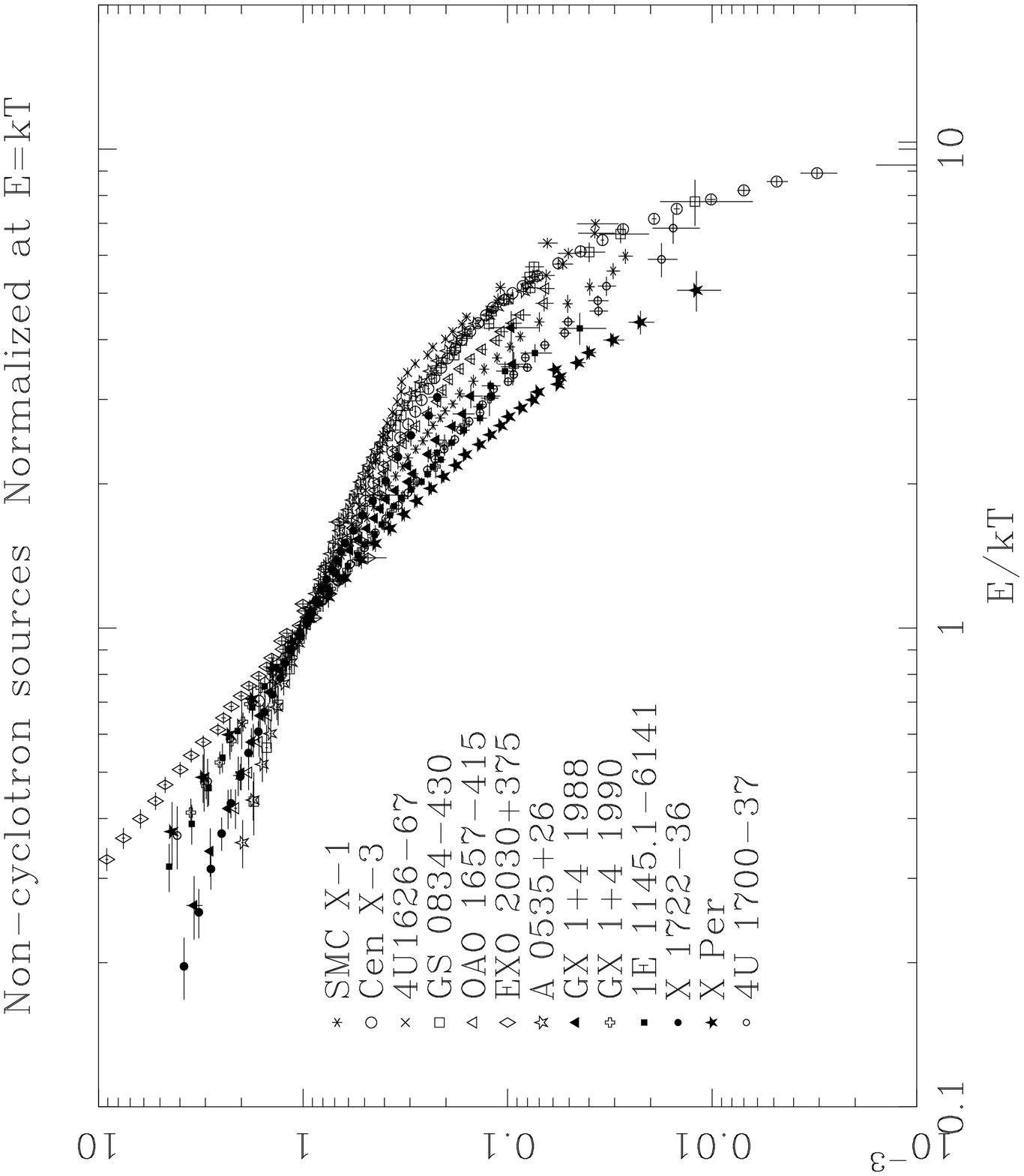}
%
\psbox[xsize=0.375#1,ysize=0.375#1,rotate=r]{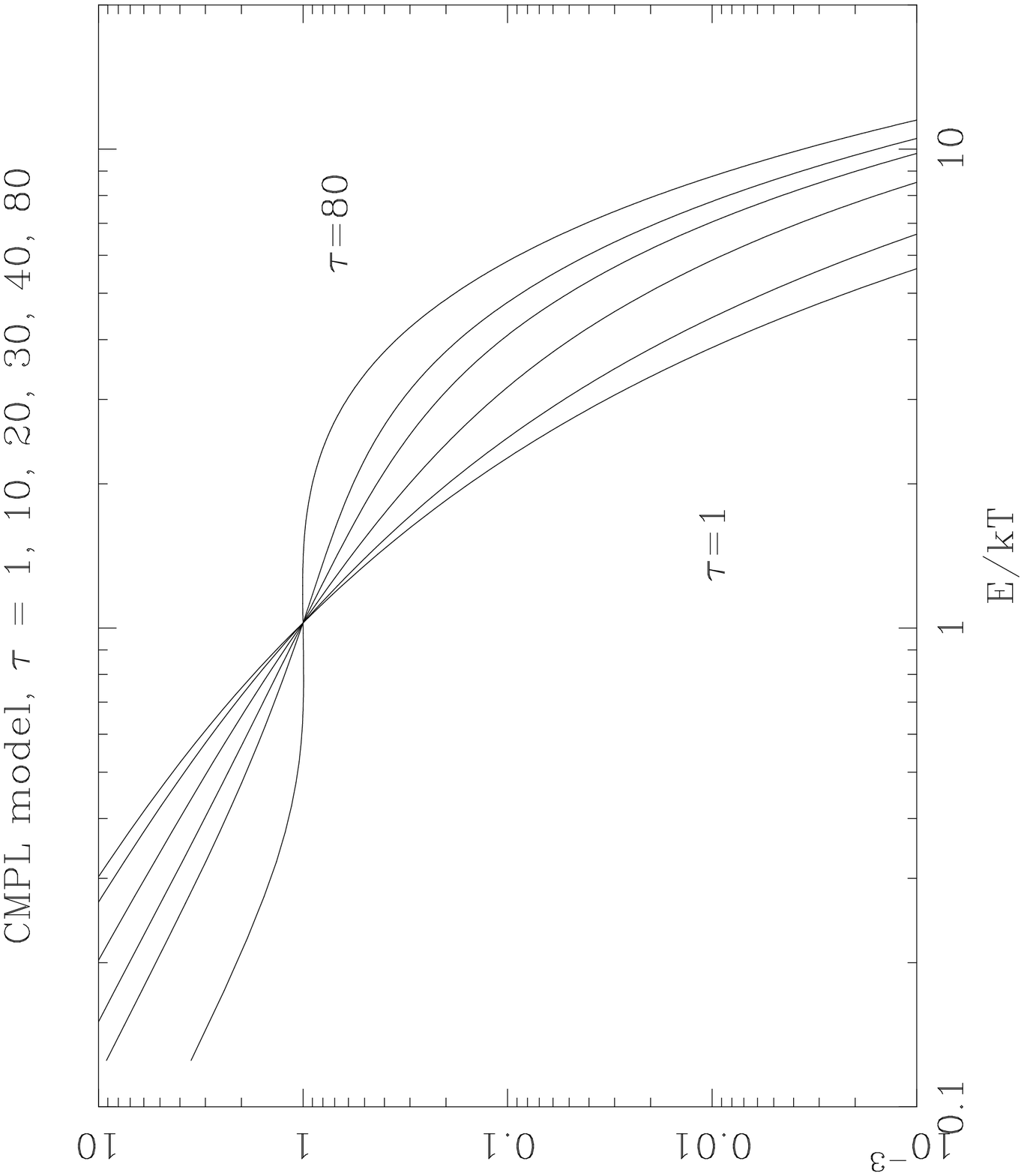}
\end{flushleft}

\vspace*{-15.9cm}

\begin{flushright}
\psbox[xsize=0.375#1,ysize=0.375#1,rotate=r]{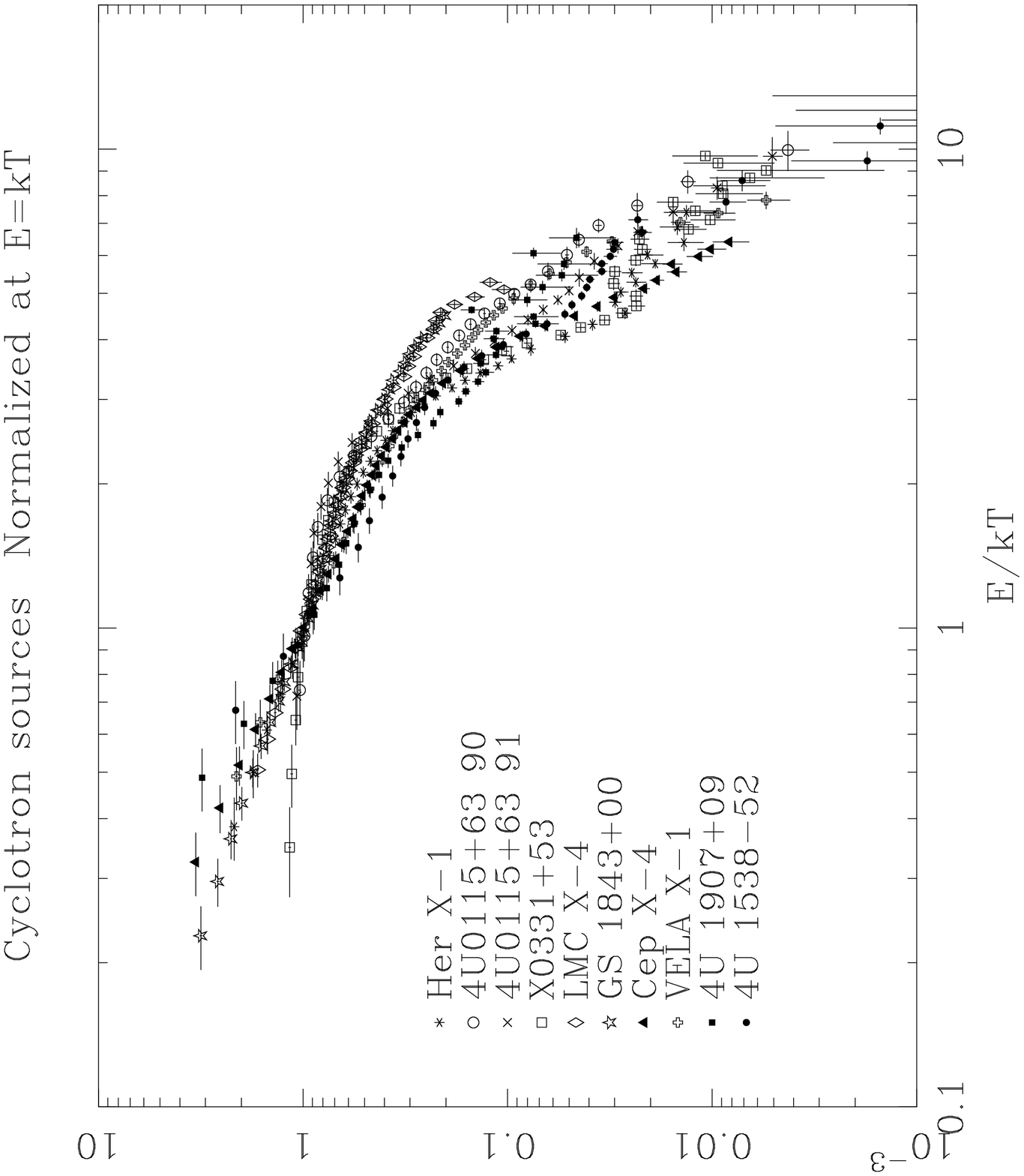}
\end{flushright}
\end{figure}
\vspace*{-0.5cm}
%

\begin{flushright}
\begin{minipage}{8cm}
\small
Fig.\ 1.\ The  continuum spectra of the non-cyclotron ({\it upper left}) and cyclotron ({\it upper right})
sources normalized by the energy $kT$ and by the flux at $E = kT$.
The instrument efficiency and $N_H$ absorption are corrected.
1E2259+586 is excluded from this figure, because it might be  contaminated by the SNR in the soft energy.
GX 301-2 is also excluded, because it has a leaky absorber.
{\it Bottom} is the Comptonization model by Lamb and Sanford (1979).
The spectral changes with the optical depth $\tau$ are very similar to the observed changes.
\normalsize

\end{minipage}
\end{flushright}

\vspace{5.5mm} \hspace{-8mm} 
Then what determines those parameters ?
The obtained $\tau$ has a negative relation with $kT$. As $\tau$ becomes thick, $kT$ goes down,
while $kT$ does not depend on $L_X$ nor spin period.
The only parameter which has a possible relation with $kT$ is the resonance energy $E_a$ (Figure 2).
In the plasma where scattering is dominant, 
the energy transfer from an electron to a photon is  given by the
Kompaneets equation (eg.\ Rybicki and Lightman 1979),
as $ \Delta E = E/(mc^2) (4kT - E)$.
In an equilibrium,
$ E = 4kT $.
Now the interacting photons are mainly that with the resonance energy because the cross section is extremely large.
Consequently the temperature of the electrons is `adjusted' to satisfy equation $ kT = 0.25 E_a $.
Monte-Carlo simulations of optically thick media by Lamb \etal (1990) find $kT \approx 0.27 E_a$,
and it is applied to the $\gamma$-ray bursts.
The $kT = 0.25 E_a$ relation, shown in Figure 2, is in a rough agreement with the data points.

\vspace{5.5mm} \hspace{-8mm} 
Let us make sure that the Comptonization is the dominant process in the accretion column.
Protons have most of the gravitational energy in the accreting matter.
The time scales in which protons give energy to electrons, $t_{col}$, and electrons lose energy by the Comptonization, $t_{comp}$, are
\[
t_{col} = 5 \times 10^{-5} \ n_{20}^{-1} \ \biggl(\frac{kT}{\mbox{10 keV}}\biggr) \ \ \mbox{[s]}, \hspace{1cm}
t_{comp} = 1 \times 10^{-15} \ \biggl(\frac{kT}{\mbox{10 keV}}\biggr)^{-4} \ \ \mbox{[s]}
\]

\begin{figure}[htb]
\begin{flushright}
\psbox[xsize=0.4#1,ysize=0.4#1,rotate=r]{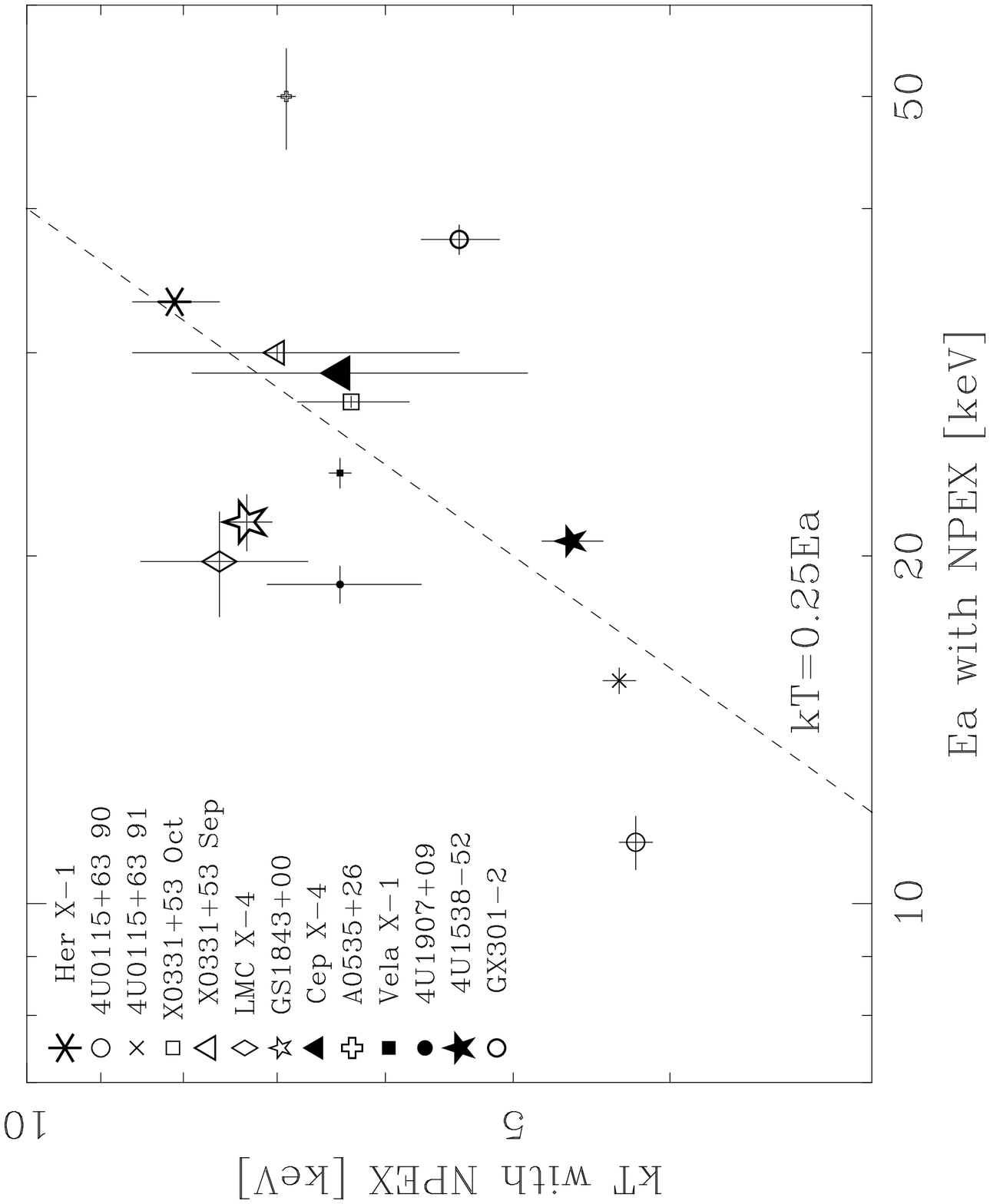}
\end{flushright}
\end{figure}

\begin{flushright}
\begin{minipage}{8.5cm}
\vspace{-10mm}
\small
Fig.\ 2.\ Correlation between  $E_a$  and $kT$. There is a weak positive relation.
If the Comptonization is dominant, $ kT \approx 0.25 E_a $ is expected.
\normalsize
\end{minipage}
\end{flushright}
\vspace{-108mm}

\begin{flushleft}
\begin{minipage}{8cm}

(Gould 1982, Rybicki and Lightman 1979). Here $n_{20}$ is the density in the unit of $10^{20}$ cm$^{-3}$, and
Compton cross section is assumed to be $10^4 \sigma_T$  near the cyclotron resonance.
Therefore electrons and photons interact much more strongly than protons and electrons, and
electrons and photons are in the Comptonization equilibrium.

\vspace{5.5mm}
From the $E_a-kT$ relation  one important suggestion can be deduced.
Pulsars have $kT$ between 4--14 keV (Table 1),
which might indicate the cyclotron resonance energies are fairly constant within 10-60 keV.
Moreover, the temperatures of the non-cyclotron sources are relatively higher than those of the cyclotron sources,
which might mean that possible resonances are nearly at the high end of the energy range of \ginga
and they are difficult to detected.

\end{minipage}
\end{flushleft}

\vspace{3mm} \hspace{-8mm} 
What is the source of the input soft photons, then ?
The bottom of the accretion column or the neutron star surface are candidates.
From the observational view, Her X-1 has a strong soft 0.1keV blackbody component (McCray 1982).
Although its origin is said to be the inner accretion disk or the Alfven shell, 
some of it might come directly from the bottom of the accretion column.

\vspace{5.5mm} \hspace{-8mm} 
We have used the Comptonization model without magnetic fields.
Although the scattering cross section of an electron heavily depends
on its energy in the magnetic fields, 
Meszaros (1992) notices that the continuum spectrum would be similar even in the magnetic fields
except for the resonance.
An absorption or an emission feature would be formed at around the resonance depending on the geometry and
the optical depth of the scattering plasma.
Readers might feel as if the Comptonized continuum is absorbed by CYAB,
but it is not true.  Those two are formed at the same time by the same scattering process.

\section*{\normalsize MAGNETIC FIELDS DISTRIBUTION}

We found the cyclotron structures from 11 pulsars among 
23 X-ray pulsars including 1 pulsar-candidates.
Adding A0535+26 from which {\it HEXE} discovered the
cyclotron line (Kendziorra 1994), the cyclotron structures were detected from 12
pulsars, about a half of the 23 sources.
Now we can make a distribution of the magnetic fields (Figure 3).
The magnetic fields range between $3 \times 10^{11} - 5 \times 10^{12}$ G,
which is similar to the life-corrected distribution  of the radio pulsars ({\it right dotted line}),
ranging between  $10^{11} - 10^{13}$ G with a peak at $2 \times 10^{12}$ G.
The distribution of X-ray pulsars looks different from
that of the {\it observed} radio pulsars ({\it right solid line}),
which might indicate that the magnetic fields of the X-ray pulsars do not decay within a characteristic  time scale of the radio
pulsars ($10^{6}-10^{7}$ y).
As the magnetic fields of the radio pulsars are obtained assuming the magnetic dipole radiation, only
the dipole component are measured.
On the other hand, those measured by the X-ray cyclotron structure are almost on the surface of the neutron star
and contain all multipole components.
The agreement of the two indicates that the magnetic fields of the neutron star is dipole, and not multipole.

\begin{figure}[thb]
\vspace{-6mm}
\begin{flushleft}
\psbox[xsize=0.43#1,ysize=0.43#1,rotate=r]{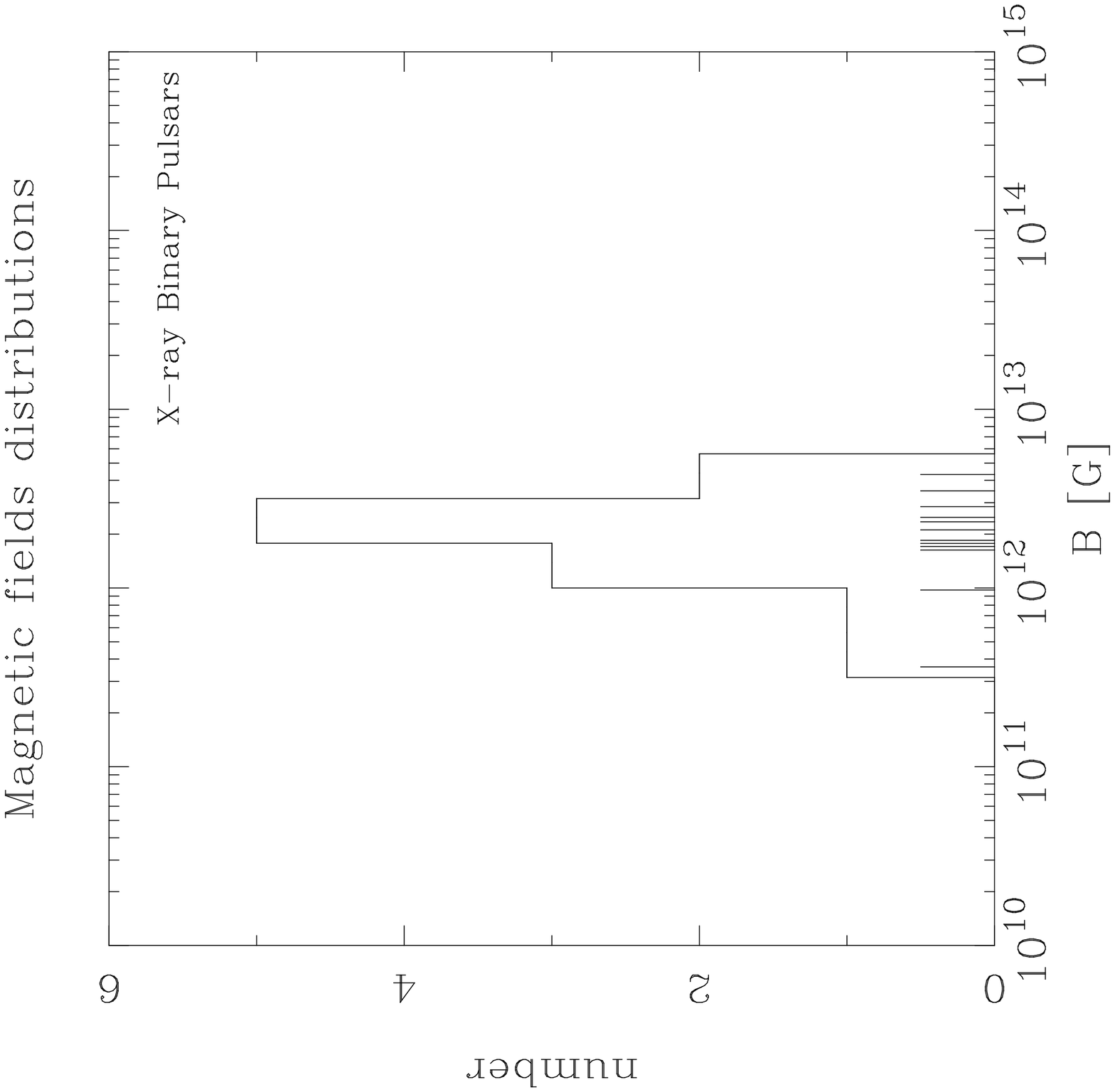}
\end{flushleft}
\vspace{-5mm}
\small
Fig.\ 3.\ The magnetic fields distribution of X-ray binary pulsars measured by the cyclotron resonances ({\it left}).
The magnetic fields of the 12 pulsars are indicated with short lines on the horizontal axis.
The distribution is similar to the life-corrected distribution of the radio pulsars ({\it right dotted line}),
ranging in $10^{11}-10^{13}$ G with a peak at $2 \times 10^{12}$ G.
The {\it right solid line} is that of the observed radio pulsars.
\vspace{3mm}
\normalsize
\end{figure}

\vspace{5.5mm} \hspace{-8mm} 
Let us discuss the selection effect. If there is a pulsar with a
cyclotron resonance of less than a few keV,
it is expected to show a steep power-law spectrum ($\alpha \sim 3$) 
in the \ginga energy range, as  1E2259+586.
But all the other pulsars show a flat power-law in 2--10 keV, which 
suggests that pulsars with $E_a$ $^{\displaystyle <}_{\displaystyle \sim} 2 $ keV
does not exist.
Although the detection limit towards the high energy  is due to
the \ginga energy range,
$kT$ and the $E_a-kT$ relation predict all the pulsars would have the cyclotron resonances in 4--60 keV.
The pulsars with $E_a > 50$ keV are not likely.
Since the data in Figure 3 already contain half the sources in the class, 
eventual inclusion of the others, even if all are at higher and lower 
energies, will not much change the distribution function of Figure 3.  
Thus, the magnetic field of X-ray binary pulsars are likely to cluster between  $3 \times 10^{11} - 5 \times 10^{12}$ G.

\section*{\normalsize RESONANCE ENERGY CHANGE}

As shown in Figure 4 the cyclotron structure of 4U 0115+63 changed between the two observations.
In 1990 it had double harmonic structures with the fundamental resonance at $E_a = 11$ keV.
In 1991, however, it showed a broad single structure centered at $E_a = 16$ keV.
It showed double/single structure throughout the pulse phases in 1990/1991, respectively.
The luminosity in 1991  was 1/6 of that in 1990.
\ginga observations of some sources with different intensities are
summarized in Table \ref{Ea intensity}.
If we tentatively attribute the change of $E_a$ to the height change of the scattering region ($E_a \propto r^{-3}$)
and calculate the height difference assuming $r$ in the weaker state
is equal to the radius of the neutron star $R_{NS}$ = 10 km,
the height change is  as much as 1.1 km in 4U 0115+63 as listed in Table 2
$\Delta$height column.

\vspace{5.5mm} \hspace{-8mm} 
Let us estimate the height of the accretion column employing the model by  Burnard \etal (1991)
to examine whether the change in height cited in Table \ref{Ea intensity} is reasonable or not. 
In the case of a pulsar, the accretion stream concentrates on the magnetic poles.
Therefore Eddington limit of the emission along the magnetic fields is only $10^{35.7}$ erg/s.
However, if the emission is sideward, most of the photon pressure is supported by the magnetic fields without stopping the
accreting matter. Then the `Eddington limit' $L_1$ becomes
\[
L_1 \ = \theta_c L_{Edd} H_\perp \sim \ 10^{37.3} \ \frac{\theta_c}{0.1} \ \biggl(\frac{M_{NS}}{1.4 M_\odot}\biggr) \ H_\perp \  \mbox{  ergs/s} .
\]
Here $\theta_c$ ($\sim 0.1$) is the opening angle of the accretion column, $H_\perp$ ($\sim 1.3$)
is the ratio of the Thomson cross section and
the Rosseland averaged cross section for the radiation flow across $B$.
$L_{Edd}$ ( = $2.0\times 10^{38}$ erg/s) is the conventional Eddington Luminosity by the Thomson scattering.

\vspace{5.5mm} \hspace{-8mm} 
When a pulsar emits as much as $L_1$,
accretion flow yields 
a mound on the surface, whose
height $H_s$ would change 
in proportion to $L_X$. 
\begin{equation}
\label{HtoLnoshiki}
 H_s \approx \frac{L_X}{L_1} R_{NS} \theta_c = \frac{L_X}{L_{Edd} H_\perp}  R_{NS}
\end{equation}
Detailed calculations by Burnard \etal (1991) and Basko and Sunyaev (1976) justify this relation
showing that the height $H_s$ is the place where the radiation-dominated shock
at Thomson optical depth $\sim 4-9$ transforms
free fall matter into the subsonically settling on the mound.

\vspace{5.5mm} \hspace{-8mm} 
We choose $kT$ of the NPEX model as the temperature, assume $R_{NS}$ = 10 km and $M_{NS}$ = 1.4 \solarmass, and
calculate $H_\parallel$ and $H_s$ as listed in Table \ref{Ea intensity}.
$E_a - H_s$ relations are shown in Figure 5.
If we assume a dipole magnetic field ($B \propto r^{-3}$) and the
gravitational redshift ($E_a \propto r^{0.34}$ in $r = 10 \sim 11$ km),
the predicted $H_s$ agree very well with the observations for X0331+53
and \mbox{4U 0115+63} as indicated by the dashed lines.
The values of \mbox{4U 1538-52} are also consistent, since 
the low luminosity forms a low mound where the height change is
negligibly small
when the luminosity changes by 1.3.
The measured magnetic field of 4U 1538-52 would be almost that on the surface of the neutron star.
These very good agreements of the observations and predictions
support the assumption that the change of the resonance energy is caused by the height change of the accretion column
by the luminosity and that the magnetic field is dipole. Some
questions, however,  remain such as why the cyclotron scattering is
dominated at the top of
the accretion column while the most emission is from the bottom of it.

\vspace{5.5mm} \hspace{-8mm} 
Her X-1 does not appear to obey the relation, and $E_a$ is changing independently of $L_X$.
However, it has the 35 d intensity cycle and there are many reasons to change the apparent luminosity of Her X-1,
such as an increase of the scattering gas, occultation by the accretion disk,
change of the X-ray beam  and so on.
The three points of $E_a$ in 1989 were on one line, and a point in 1990 is off.
The circumstances might not have changed much during the same or sequent main-on.

\vspace{5.5mm} \hspace{-8mm} 
In Cep X-4 we cannot calculate $H_s$ since we do not know the distance to it.
But by assuming the relations (eq.\ \ref{HtoLnoshiki}) and $E_a \propto r^{-2.66}$, 
we can obtain the distance.
Unknown parameters are the distance and the surface magnetic
field and we have three data points to be fitted.
We obtain the distance to Cep X-4 of 3.2 $\pm$ 0.4 kpc.
The $H_s$  are calculated to be  210 m, 170 m, and 110 m  on 1988/4/3, 
8, 14, respectively.
The luminosities are log$_{10} L_X$ = 36.75, 36.67, and 36.48, respectively.
This can be a new method to estimate the distance to a binary X-ray pulsar.

\vspace{5.5mm} \hspace{-8mm} 
\small
NOTE: Cep X-4 was optically identified (Bonnet-Bidaud 1997, {\it IAU Circ.} 6724) using
the position by ROSAT from the 1993 outburst (Schulz 1995, \aaa, {\bf 295}, 413).
The distance is 2.3-2.7 kpc from the reddening assuming
the density of 1 H-atom cm$^{-3}$, or 3 kpc from the 
strong Na absorption line. 
Those are roughly consistent with our result.
\normalsize

\vspace*{3mm}

\begin{figure}[htb]
\psbox[xsize=0.65#1,ysize=0.65#1,rotate=r]{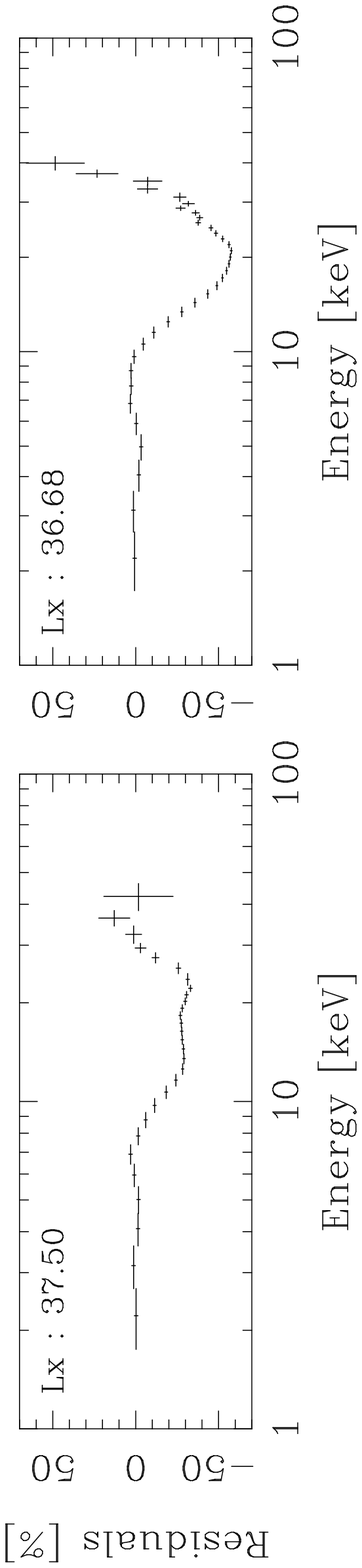}
\end{figure}

\vspace*{-9.1cm}
\hspace{-8mm} 
\small
Fig.\ 4.\ Residuals from simple NPEX fits of 4U 0115+63 in 1990 ({\it left}) and 1991 ({\it right}).
The cyclotron structure changed dramatically in the shape and the depth. 
The NPEX parameters are listed in Table 1.
\normalsize

\vspace{2mm}

\vspace{1mm}
\section*{\normalsize REFERENCES}
\vspace{1mm}
\small
\begin{list}{}{\setlength{\leftmargin}{3em}\setlength{\rightmargin}{0cm}
	\setlength{\itemsep}{0.0ex}\setlength{\baselineskip}{0ex}
	\setlength{\itemindent}{-3em}}
\item Arons, J., R. I. Klein, and S. M. Lea, {\apj}, {\bf 312}, 666 (1987). 
\item Basko, M. M., and R. A. Sunyaev,   \mnras, {\bf 175}, 395 (1976). 
\item Burnard, D. J., J. Arons, and R. I. Klein,   {\apj}, {\bf 367}, 575 (1991).  
\item Gould, R. J.,  Processes in Relativistic Plasmas, {\apj}, {\bf 254}, 755 (1982). 
\item Kendziorra, E., P. Kretschmar, H. C. Pan, M. Kuntz, M. Maisack \etal,  \aaa, {\bf 291}, L31 (1994). 
\item Lamb, P. and P. W. Sanford,     \mnras, {\bf 188}, 555 (1979). 
\item Lamb, D. Q., C. L. Wang,  and I. M. Wasserman,  \apj, {\bf 363}, 670 (1990). 
\item Makishima, K., and  T. Mihara,  Magnetic Fields of Neutron Stars,
      {\it Frontiers of X-ray Astronomy}, p23, ed. Y. Tanaka, and K. Koyama,
	  Universal Academic Press Inc., Tokyo (1992).	
\item McCray, R. A., J. M. Shull, P. E. Boynton, J. E. Deeter, S. S. Holt, \etal, 
EINSTEIN Observatory Pulse-phase Spectroscopy of Hercules X-1, 
\apj, {\bf 262}, 301 (1982).	
\item Meszaros, P.,  {\it High-Energy Radiation from Magnetized Neutron Stars}, University of Chicago Press (1992).
\item Mihara, T., K. Makishima, T. Ohashi, T. Sakao, M. Tashiro \etal,  \nature, {\bf 346}, 250 (1990).	
\item Mihara T.,  Ph.D. thesis for the physics degree of University of Tokyo (1995).
\item Nagase, F., Accretion-Powered X-Ray Pulsars, \pasj, {\bf 41}, 1 (1989).	
\item Nagase, F., T. Dotani, Y. Tanaka, K. Makishima, T. Mihara  \etal,  \apj, {\bf 375}, L49 (1991).	
\item Rybicki, G. R. and Lightman, A. P.,  {\it Radiative Processes in Astrophysics}, John Wiley \& Sons, Inc. (1979).
\item Soong, Y., D. E. Gruber, L. E. Peterson, and R. E. Rothschild,  \apj, {\bf 348}, 641 (1990).	
\item Sunyaev, R. A. and Titarchuk, L. G.,  \aaa, {\bf 86}, 121 (1980).	
\item Tr\"{u}mper, J., W. Pietsch, C. Reppin, W. Voges, R. Staubert \etal,  \apj, {\bf 219}, L105 (1978).	
\item Turner, M. J. L., H. D. Thomas, B. E. Patchett, D. H. Reading, K. Makishima  \etal, 
The Large Area Counter on Ginga, \pasj, {\bf 41}, 345 (1989).	
\item Wheaton, W. A., J. P. Doty, F. A. Primini, B. A. Cooke, C. A. Dobson  \etal,  \nature, {\bf 282}, 240 (1979).	
\end{list}

\begin{table}[htb]
\small
\caption[]{\small Cyclotron resonance energies with luminosities.
$H_\parallel$ is a function of $E_a/kT$ and obtained from Arons \etal (1987).
$H_s$ is the height of the accretion column calculated by eq.\ (\ref{HtoLnoshiki})
from $L_X$ and $H_\parallel$ assuming $M_{NS}$ = 1.4 \solarmass, $R_{NS} = 10$ km, and $\theta_c = 0.1$.
}
\label{Ea intensity}
\begin{center}
\begin{tabular}{ccccccccc}
\hline
\hline
\multicolumn{2}{l}{sources}& count rate & $E_a$ &  $kT$    &  log$_{10} L_X$ & $\Delta$height   & $H_\parallel$, $H_\perp$ & $H_s$ \\
& date      &  [c/s]     &     [keV]      &     [keV]       &   [erg/s]       &       [m]       &                          & [m] \\
\hline                                                                                 
\multicolumn{2}{l}{4U 0115+63} & \multicolumn{2}{l}{3--50 keV} \\
& 1990/2/11 &  4036      &$ 11.3 \pm 0.6 $&$ 4.25 \pm 0.10 $&  37.50          &$ 1100 \pm 220 $ & 1.23 & 1280 \\
& 1991/4/26 &   661      &$ 15.6 \pm 0.4 $&$ 4.34 \pm 0.14 $&  36.68          &     0           &      &  203 \\
\multicolumn{2}{l}{X0331+53} & \multicolumn{2}{l}{3--37 keV} \\
& 1989/10/1 &  3586      &$ 27.2 \pm 0.3 $&$ 6.3 \pm 0.5   $&  37.43          &$  330 \pm 70  $ & 1.44 & 930 \\
& 1989/9/20 &  2271      &$ 30.0 \pm 0.5 $&$ 7.0 \pm 1.6   $&  37.29          &      0          &      & 674 \\
\multicolumn{2}{l}{Cep X-4} & \multicolumn{2}{l}{2--37 keV} \\
& 1988/4/3  &   834      &$ 28.58 \pm 0.5                        $&$ 7.5 \pm 3.8   $&  36.75$^{b}$    &$ 82 \pm  109 $ & 1.38 & 210$^{b}$ \\
&           &            & \multicolumn{1}{r}{$ (\pm 0.05)^{a} $} &                 &                 & \multicolumn{1}{r}{$(\pm 14)$\ \ \ } & \\
& 1988/4/8  &   692      &$ 28.94 \pm 0.4                        $&$ 7.1 \pm 2.2   $&  36.67$^{b}$    &$ 40 \pm  102 $ &      & 170$^{b}$ \\
&           &            & \multicolumn{1}{r}{$ (\pm 0.05)^{a} $} &                 &                 & \multicolumn{1}{r}{$(\pm 14)$\ \ \ } &  \\
& 1988/4/14 &   450      &$ 29.29 \pm 0.8                        $&$ 6.4 \pm 1.7   $&  36.48$^{b}$    &     0          &      & 110$^{b}$ \\
&           &            & \multicolumn{1}{r}{$ (\pm 0.11)^{a} $} &                 &                 &                &      & \\
\multicolumn{2}{l}{Her X-1} & \multicolumn{2}{l}{3--60 keV} \\
& 1990/7/27 &  1154      &$ 34.1 \pm 0.4 $&$  8.1 \pm 1.2  $&  37.53          &$ -160 \pm 140 $ & 1.35 & 1250 \\
& 1989/6/3  &   857      &$ 32.5 \pm 1.0 $&$ 10.3 \pm 3.2  $&  37.44          &$   0  \pm 110 $ &      & 1020 \\
& 1989/6/6  &   792      &$ 32.5 \pm 0.4 $&$  8.2 \pm 0.9  $&  37.40          &       0         &      &  930 \\
& 1989/5/3  &   739      &$ 33.9 \pm 1.2 $&$  9.7 \pm 5.2  $&  37.37          &$ -140 \pm 123 $ &      &  860 \\
\multicolumn{2}{l}{4U 1538-52} & \multicolumn{2}{l}{3--37 keV} \\
& 1988/3/2  &   184      &$ 20.6 \pm 0.2 $&$  4.7 \pm 0.3  $&  36.56          &$  0 \pm 50    $ & 1.47 & 120 \\
& 1990/7/27 &   130      &$ 20.6 \pm 0.2 $&$  4.6 \pm 0.2  $&  36.43          &      0          &      &  91 \\
\hline
\end{tabular}
\end{center}

a:\ Single parameter error, when other parameters than $E_a$ are fixed to their best-fit values.

b:\ Estimated in this work assuming that the observed height changes are
equal to the $H_s$ changes.
They would have very large errors because of the large errors of $E_a$.
The most probable distance to \mbox{Cep X-4} is 3.2 kpc.
\end{table}

\clearpage
\begin{figure}[htb]
\begin{center}
\psbox[xsize=0.67#1,ysize=0.67#1,rotate=n]{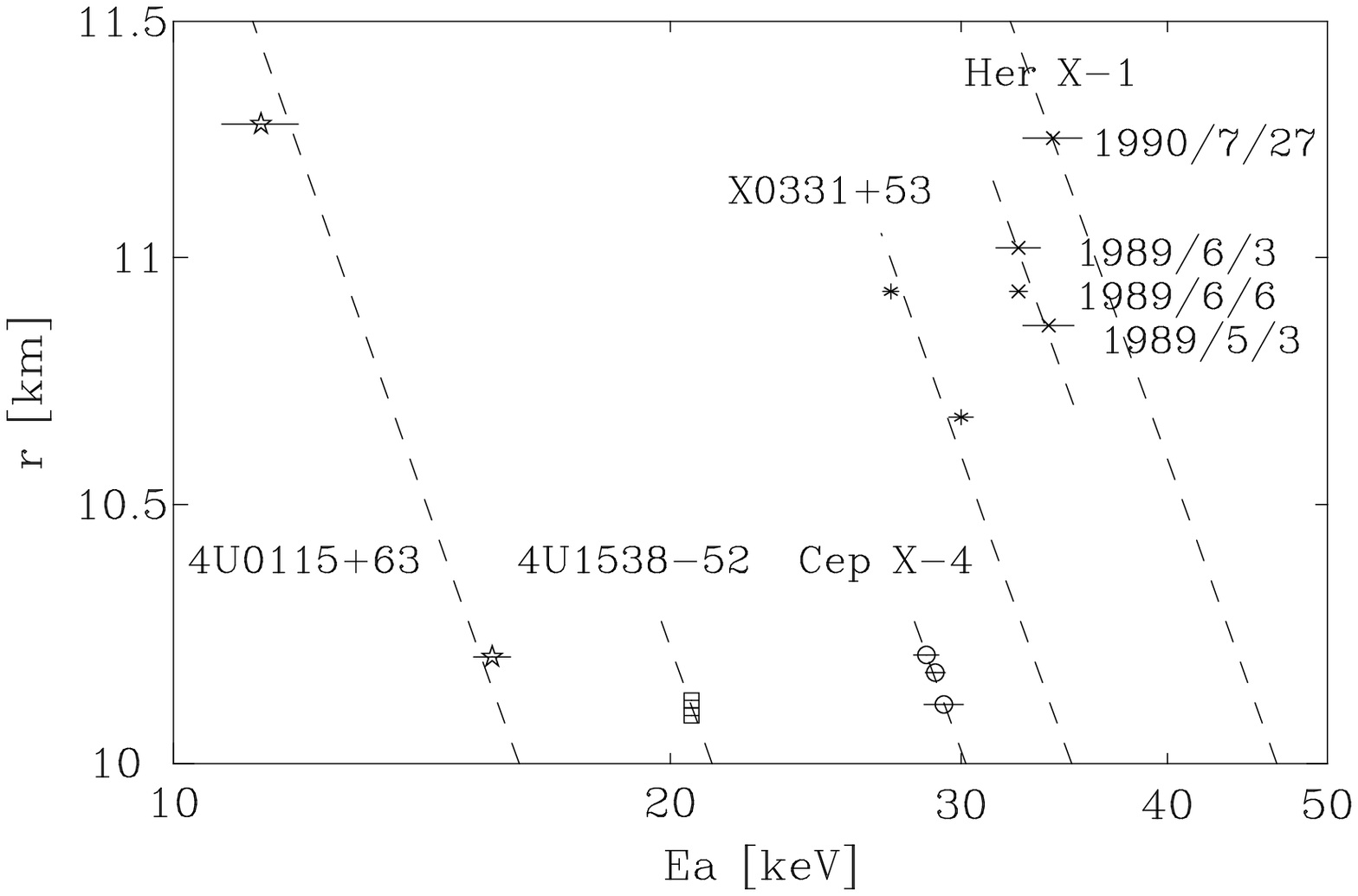}
\end{center}
\end{figure}
\vspace*{-89mm} 

\hspace{-8mm} 
\small
Fig.\ 5.\ The observed resonance energies and the heights of the
accretion column estimated by a simple theory (eq.\ 1).
$r$ is the height from the center of the neutron star.
$R_{NS}$ = 10 km and $M_{NS} = 1.4$\solarmass are assumed.
The dashed line indicates $r^{-2.66}$ dependencies of the dipole
magnetic field and the gravitational redshift.
4U0115+63, 4U1538-52, and X0331+53 obey this simple law well.
Her X-1 does not obey this law, which would have other mechanisms to
change the apparent luminosity.
Assuming the $r^{-2.66}$ relation, the luminosities of Cep X-4 are calculated,  
which leads the distance to be 3.2 kpc.
\normalsize

\end{document}